\documentclass[12pt]{spieman}  
\usepackage{amsmath,amsfonts,amssymb}
\usepackage{graphicx}
\usepackage{setspace}
\usepackage{tocloft}
\usepackage{lineno}

\title{Initial Assessment of Monocrystalline Silicon Solar Cells as Large-Area Sensors for Precise Flux Calibration}

\author[a,*]{Sasha Brownsberger} 
\author[a]{Nicholas Mondrik} 
\author[a,b]{Christopher W. Stubbs} 
\affil[a]{Department of Physics  \\
Harvard University \\
17 Oxford Street
Cambridge MA 02138, USA}
\affil[b]{Department of Astronomy \\
Harvard University \\
60 Garden St
Cambridge MA 02138, USA}

\cftpagenumbersoff{figure}
\cftpagenumbersoff{table} 
\begin{document} 
\maketitle

\begin{abstract}
As the precision frontier of large-area survey astrophysics advances towards the one millimagnitude level, flux calibration of astronomical instrumentation remains an ongoing challenge. We describe initial testing of silicon solar cells as large-aperture precise calibration photodiodes. We present measurements of dark current, linearity, frequency response, spatial response uniformity, and noise characteristics of the Sunpower C60 solar cells, an interdigitated back-contact 125mm x 125mm monocrystalline solar cell. We find that these devices hold considerable promise as large-area flux calibration sensors and warrant further characterization. 
\end{abstract}


{\noindent \footnotesize\textbf{*}Sasha Brownsberger,  \linkable{sashabrownsberger@g.harvard.edu} }

\begin{spacing}{2}   

\section{Introduction} 
\label{sec:intro}

Flux calibration remains a primary source of systematic uncertainty in the use of type Ia supernovae as probes of the history of cosmic expansion \cite{Bohlin14, stubbs2015precise,scolnic2017relative, scolnic2019next}.  One approach for flux calibration is to invoke models of photon emission spectra vs. wavelength for a simple stellar atmosphere, white dwarf stars being the most popular \cite{narayan2019subpercent}. 
An alternative approach is to use well-characterized sensors as the metrology standard for relative flux determination \cite{stubbs2006toward, bernstein2017instrumental, lombardo2017scala,stubbs2010precise, Regnault15}. In this approach a well-calibrated photodetector, known to better than a part per thousand, \cite{NIST1998spectroradiometric} is used to map out the instrument's relative sensitivity vs. wavelength. 
Conventional photon-detectors (photodiodes, CCDs, etc) have collection areas no larger than a few square centimeters.  Such small collection areas are inadequate for those modern imaging applications that depend on the calibration of a large-diameter optical beam.  

The Large Synoptic Survey Telescope \cite{LSST09} (LSST) project intends to use a collimated, monochromatic beam (a Collimated Beam Projector, or CBP), to sequentially illuminate portions of the optics, and a calibrated silicon photodiode to monitor the flux \cite{ingraham2016lsst, coughlin2016collimated}.  The LSST team plans to use the CBP to measure instrument transmission as a function of photon wavelength and source position.  

One metrology challenge of this approach is to measure the flux emanating from the CBP, which has an exit pupil diameter of 240 mm, about twenty times larger than the diameter of typical silicon photodiodes.  The LSST team is considering several solutions to this unsolved problem, including changing the projector beam focus to reduce the spot size, collecting the light with a focusing concentrator, using an integrating sphere with a 250mm port, or scanning the exit beam across one or more standard calibrated photodiodes.  Each of these approaches, all suboptimal, are born of a perceived need to measure a large light source with a small detector.  

There is nothing fundamental to the process of flux calibration that mandates the use of a sensor with a small collecting area. Indeed, photon-to-electron converters with large collection areas are already in use in the form of photovoltaic cells.  Using a large-area calibration sensor allows for increased photon detection rates, without the need to use intervening focusing optics that themselves require calibration.  Additionally, modulation of the calibration light can be used to discriminate between background and calibration illumination.  In this paper we consider the possibility of using an array of high-efficiency solar cells as a full-aperture sensor for calibrating the LSST CBP and other large-diameter light sources.

In Section \ref{sec:backContact}, we describe the general architecture of an interdigitated back-contact monocrystalline silicon solar cell and introduce the C60 solar cells that we study.  In Section \ref{sec:characterization}, we describe the methods and results of our measurements.  
We conclude in Section \ref{sec:conclusion}, where we summarize the results of our experiments and discuss what additional steps should be taken to 
fully assess the prospect of creating a precision photometric calibrator composed of solar cells. 

\section{Back-Contact Silicon Solar Cells} \label{sec:backContact} 

As with astronomical CCDs and other silicon photosensors, solar cells exploit the band-gap in silicon to convert incident photons into electron-hole pairs. The challenge is to extract these charge carriers from the bulk material. 

We seek a two-terminal device that sweeps charge carriers out of the bulk and into an external circuit, where photocurrent is used to determine incident flux.  A standard approach is to apply a bias voltage along the photon incidence direction to produce an axial electric field that sweeps the charge out of the device. This requires electrical contact to the surface that the light strikes, which can compromise efficiency due to reduced fill factor or increased reflection. Back-contact solar cells\cite{Mulligan04, granek2010dissertation, liu2018review}, on the other hand, have both the anode and cathode on the non-illuminated (``back'') side of the silicon slab.  Doping on the front side serves to repel charge carriers and drive them to the anode (or emitter) where they are collected. Implants aligned with the back-side electrodes produce a series of PN junctions. 

\subsection{Conceptual Design of Back-Contact Solar Cells}
\begin{figure}
    \begin{center}
    \begin{tabular}{c} 
    \fbox{\includegraphics[width=10.0cm]{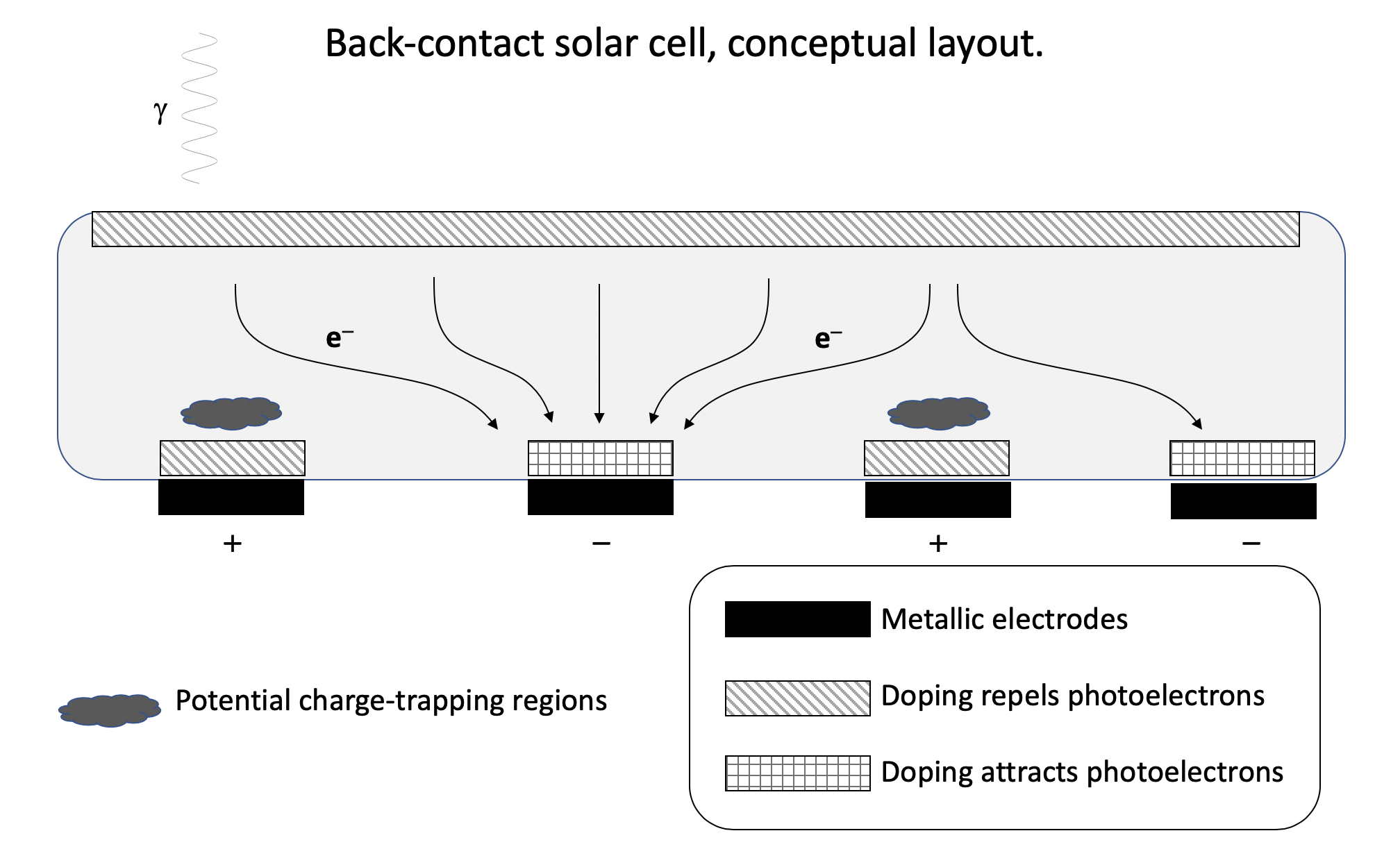}}
    \end{tabular} 
    \end{center} 
    \caption
    { \label{fig:architecture} Simplified illustration of an interdigitated back-contact solar cell architecture\cite{liu2018review}. The polarity labels match those on the C60 solar cells and indicate the direction of current flow when an external load is attached.  Light incident from the top in this diagram creates electron-hole pairs in the monocrystalline silicon, and the resulting photocharge moves to back-side electrodes that reside under doped regions. The PN junction occurs between the electrodes and creates lateral electrical fields. A layer of doping on the top surface drives the electrons away from surface states and creates a local vertical electric field component. 
    As we discuss in Section \ref{sec:spatial}, the periodicity of the electrode/doping structure may give rise to regions where charge trapping can occur, degrading the effective QE.} 
\end{figure}

We present a conceptual illustration of the back contact solar cell layout in Figure \ref{fig:architecture}.

The device's quantum efficiency (QE), the number of collected charge carriers per incident photon, is a function of incident photon wavelength, $\lambda$.  As with CCDs, the index of refraction mismatch produces reflections, especially for $\lambda <$ 400 nm.
At long wavelengths the limitation is photon energy approaching the bandgap.  

In back contact solar cells, the electrical potential is periodic in the direction perpendicular to the electrodes, creating a series of regions where the perpendicular electric field is zero.  This will reduce the QE of the device if the charge carriers dwell in these regions of null electric field long enough to recombine, or if these corrugations in the electrical potential trap photon-excited charges.  We search for and discuss the spatial dependence of a cell's effective QE in Section \ref{sec:spatial}.  

Other internal loss mechanisms can also reduce the solar cell QE.  These mechanisms include charge traps (local minima in the electrical potential energy), undesirable energy levels from contamination and defects, and electron-hole recombination. Manufacturers have worked hard to maximize the QE of silicon solar cells by minimizing these effects and by applying an anti-reflection (AR) coating to the detecting surface to minimize reflections. The QE of contemporary solar cells is in excess of 95\% over much of the wavelength range of interest \cite{Yoshikawa17}. Detailed simulations of the charge collection efficiency as a function of electrode geometry and doping concentration\cite{reichel2011investigation} have been useful in guiding the design of these devices.  

\subsection{A Useful Equivalent Circuit}\label{sec:cellModel}
\begin{figure}
    \begin{center}
    \begin{tabular}{c}
    \includegraphics[height=5.0cm]{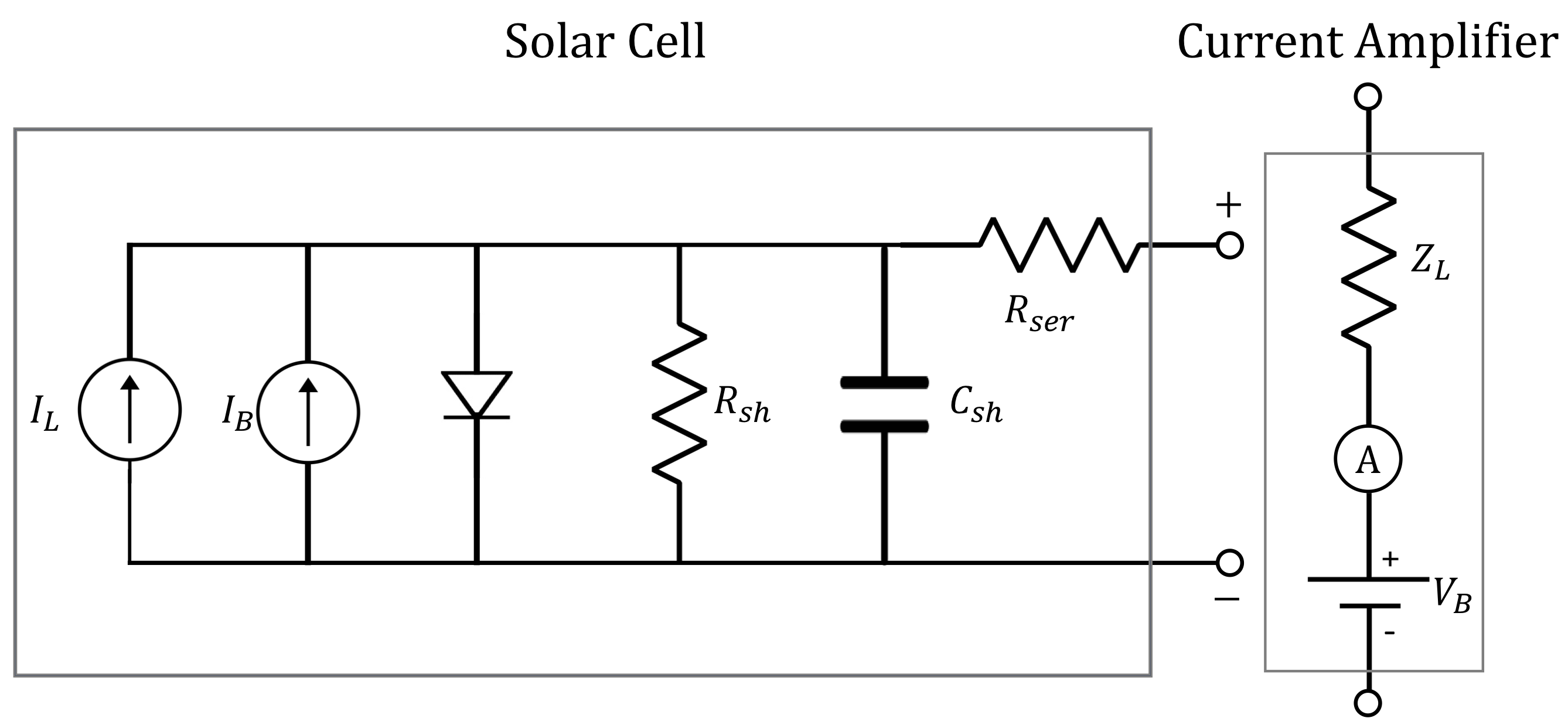} 
    \end{tabular}
    \end{center}
    \caption
    {\label{fig:equivcircuit} Equivalent Circuit for the C60 Solar Cell (left) and for the current amplifier used to measure the photocurrent (right). Our measurements suggest that every C60 solar cell has distinct characteristic values.  Typically, $R_{ser}\sim \mathcal{O}{(0.01-0.1 \Omega)}, R_{sh}^{1 \mu A} \sim \mathcal{O}{(100-1000 \Omega)}$, and $C_{sh} \sim \mathcal{O}{(1-10 \mu \mathrm{F})}$. The arrows of the current sources indicate the direction of current flow relative to the polarity designations on the electrodes of the C60 solar cell.}
\end{figure}

In analyzing the properties of a solar cell, we found it useful to model the cell's electrical behavior using a simple equivalent circuit.  In this work, we use the standard single-diode equivalent circuit \cite{Chan86} shown in Figure \ref{fig:equivcircuit}.  A more robust analysis could generalize to the two diode model \cite{Enebish93, Hovinen94} to account for distinct recombination timescales. 

In this model circuit, a pair of current sources drive current towards the output terminals.  One source, $I_L$, increases monotonically with the incident light.  The other, $I_B$, sources the bias current.  With no external load, these currents flow through the system's internal diode, shunt resistance, $R_{sh}$, and shunt capacitance, $C_{sh}$.  When the solar cell is connected to an electrical load, such as a current amplifier, current is also driven across this load in series with the cell's series resistance, $R_{ser}$. 
Our tests in Sections \ref{sec:dark} and \ref{sec:frequency} find that this single diode circuit is a good model for the C60 solar cells, with one notable exception:
when back-biased, the cell's current response deviates from the linear behavior expected from $R_{sh}$, presumably due to quantum tunneling\cite{tunnelling}. 

\subsection{A Specific Example: C60 Solar Cells}\label{sec:C60s} 

We obtained a sample of Sunpower C60 monocrystalline back-contact silicon solar cells (`C60 solar cells' henceforth) from the Amazon online storefront.  

The C60 solar cells are 125mm truncated square silicon elements with interdigitated anode and cathode electrodes. The front (illuminated) side has no electrical connection, and the electrodes on the back (dark) side have solder tab connections.  The vendor advertises an AR coating as having been applied to enhance photon conversion efficiency in the blue.  The vendor claims a QE of greater than 0.95 for $\lambda$ between 500nm and 900nm and a QE of greater than 0.9 for $\lambda$ between 400nm and 1050nm\cite{C60Brochure}. 

\begin{figure}
    \begin{center}
    \begin{tabular}{c}
    \includegraphics[height=5.5cm]{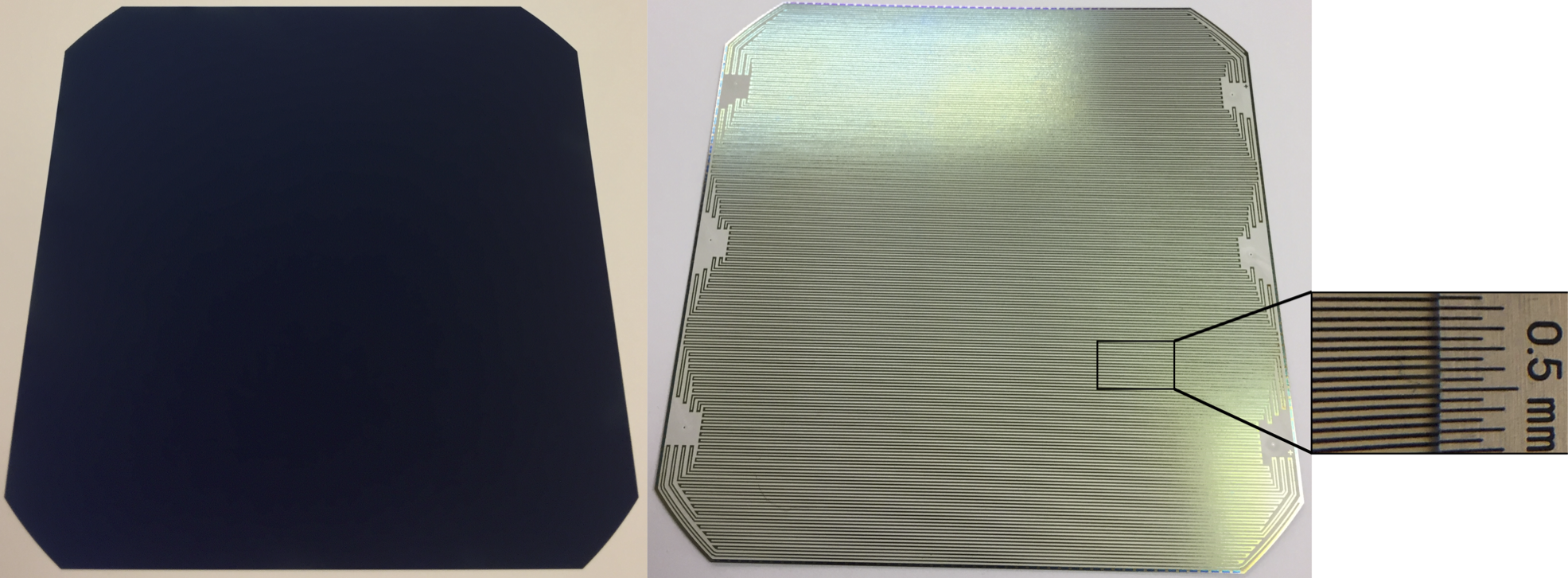}
    \end{tabular}
    \end{center}
    \caption{\label{fig:deviceimages} Front (left) and back (right) of a C60 solar cell.  The cells are 125 mm squares, with truncated corners.  The magnification of the back side shows the electrodes (silver bands) fused to the back of the silicon surface (dark bands) with a ruler for scale.  Parallel electrodes sit alternately at positive and negative voltages.  Parallel electrodes are 0.46mm apart and parallel electrodes of matching electric potential are 0.92mm apart.}
\end{figure}

The vendor's specifications are listed in Table \ref{tab:C60specs}.  Front and back images of a C60 solar cell are shown in Figure \ref{fig:deviceimages}. 

\begin{table}[htp]
    \centering
    \begin{tabular}{|l|l|}
    \hline
    Property & Value \\
    \hline
    Material & Monocrystalline silicon\\
          Side Length    &  125 mm\\
 Thickness & 165 $\mu$m\\
 Short-circuit current in 1000W/m$^2$ illumination     & 6.3 A \\
 Open circuit voltage in 1000W/m$^2$ illumination & 0.68 V \\
 Max-power output voltage in 1000W/m$^2$ illumination& 0.58 V \\
 Max-power output current in 1000W/m$^2$ illumination& 5.9 A \\
 \hline
    \end{tabular}
    \caption{Manufacturer's Characteristics of C60 Solar Cell\cite{C60Brochure}.}
    \label{tab:C60specs}
\end{table}
\begin{tabular}{c|c}
\end{tabular}

\section{Preliminary Characterization of C60 Solar Cells}\label{sec:characterization} 

\subsection{Preparing the Cells for Use} \label{sec:prep} 
The C60 solar cells do not come with any external wires for connection to other devices.  The manufacturer recommends that the user directly solder a pair of specially designed ``dogbone'' connection tabs, one for the anode and one for the cathode, to the system's solder points.  However, we found that soldering the solar cells directly caused their unbiased dark currents to increase by a factor of $5-100$ or more.  
Any non-photoelectric current introduces measurement error (both statistical and systematic), and ought to be minimized.  

We soldered wire leads to the dogbone tabs and used electrically conductive copper tape, reinforced with non-conductive super glue, to affix the dogbone tabs to the solar cells.  This method produced stable connections at room temperature without degrading the dark current.  However, we are concerned about these attachment's multi-year reliability and their stability under large current loads.  We suspect that a conducting adhesive would yield better long-term results. 

\subsection{Unilluminated IV Curve of C60 Solar Cell} \label{sec:dark} 
As we model in Figure \ref{fig:equivcircuit}, a current amplifier attached to the output solar cell imposes a bias voltage, $V_b$, and carries a load impedance, $Z_L$.  An ideal current amplifier has both $V_b=0$ and $Z_L =0$.  
When a current, $I_{Out}$, passes through the load, the voltage drop over the series resistance and the load will drive a current, $I_{sh}$, through the shunt impedance, $Z_{sh}$: 
\begin{equation}
I_{sh} Z_{sh} = V_b + I_{Out} (R_{ser} + Z_L) \ .
\end{equation} 
Current generated within the solar cell, $I_{SC}$, will split between the shunt impedance and the load: 
\begin{equation}
I_{SC} = I_{Out} + I_{sh} \ .
\end{equation} 
The fraction of the solar cell current that passes through the load is thus dependent on the applied bias voltage and the ratio of the impedances: 
\begin{equation} \label{eq:signalCurrent} 
\frac{I_{Out}}{I_{SC}} = \frac{r}{1+r} - \frac{V_b}{(R_{ser} + Z_{L} + Z_{sh}) I_{SC}} \ ,
\end{equation} 
where $r \equiv Z_{sh} / (R_{ser} + Z_L)$ defines a figure of merit for the solar cell and load.  For light sources modulated at frequencies less than 100Hz, $Z_{sh} \simeq R_{sh}$.  

\begin{figure}
    \begin{center}
    \begin{tabular}{c}
    \includegraphics[height=6.0cm]{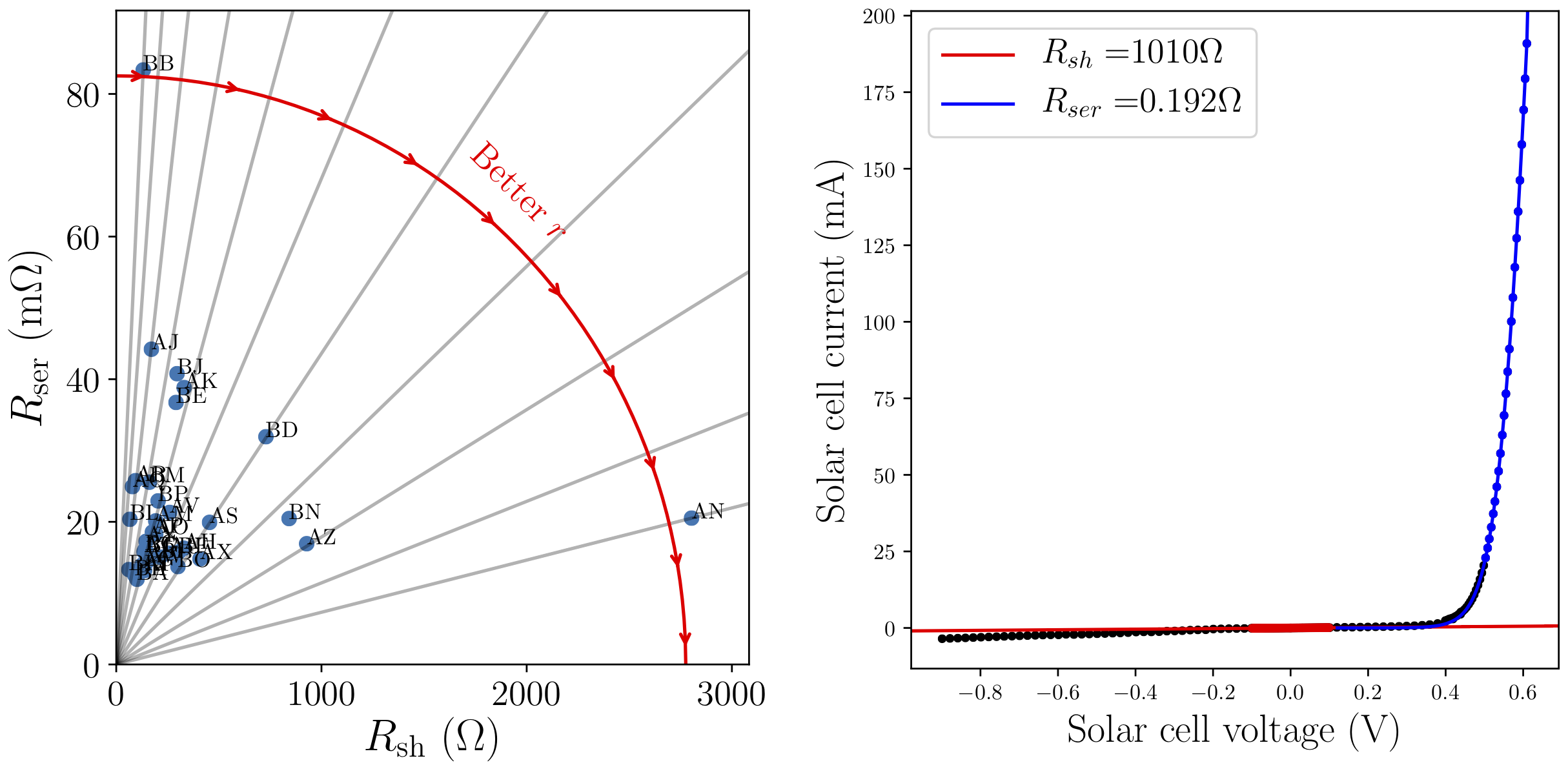}
    \end{tabular}
    \end{center}
    \caption{\label{fig:RShuntVsSeriesAndDarkIV} The measured shunt and series resistances ($R_{sh}$ and $R_{ser}$) of 37 unmodified C60 solar cells (left) and the full unilluminated IV curve of the AX solar cell after wiring.  In the left plot, each point is a different solar cell with a sequentially assigned label and the lines show increasing values of the figure of merit, $r=R_{sh}/(R_{ser}+R_L)$ when the load resistance, $R_L$, is negligible.  Those cells with the largest value of $r$, in this case cells AN, AZ, and BN, are the best candidates for use in a precision photometric calibrator.  In the right plot, we measured a shunt resistance of $1010\Omega$ and a series resistance of $0.192\Omega$.  Comparing these results to the measured resistances before the cell was wired (the point labeled ``AX'' in the left plot), the shunt resistance is about 120\% larger and the series resistance is about 1300\% larger.  The increased series resistance is likely due to the resistance of the affixed wiring, and underscores the importance of good electrical contacts and short electrical leads.  The increase in shunt resistance is somewhat surprising, and something that we do not presently understand.}
\end{figure}

A value of $I_{Out}/I_{SC}$ less than unity reduces the effective system QE of the detector over all photon wavelengths.  Any variations in the resistances or in the bias voltage will produce a wavelength-independent, systematic change in the cell's effective QE and a corresponding error in flux measurements.  Furthermore, because the shunt impedance of the solar cell is non-ohmic (see Figure \ref{fig:RShuntVsSeriesAndDarkIV}), this QE reduction depends on the incident light intensity and is a potential source of system nonlinearity.  It is paramount that $r$ be kept as large as possible by minimizing the series resistance and by maximizing the shunt resistance.   In practice, this means making good electrical connections and selecting solar cells with the largest values of $R_{sh}$.  

To better understand the device-to-device variation in $r$, we measured the shunt and series resistances of 37 C60 solar cells before affixing the conducting tabs.  

According to the equivalent circuit model of Section \ref{sec:cellModel}, $I_{Out}$ for an unilluminated solar cell with forward-bias voltage, $V_{fb}$, is determined by Ohm's Law: 
\begin{equation} \label{eq:mShunt} 
    I_{Out} =  V_{fb}/ R_{sh}  \ ,
\end{equation}
where we ignore the cell's bias current, assume $R_{sh} >> R_{ser}$, and assume that $V_{fb}$ is less than the conducting voltage of the diode.  Using Equation \ref{eq:mShunt}, we measured the shunt resistance of each cell by fitting a line to the IV curve acquired by applying DC bias voltages ranging from -100mV to 100mV in 10mV increments to the output terminals of unilluminated solar cells. 

When the applied voltage biases the diode into conduction, $R_{sh}$ is shorted and has a negligible effect on the cell's internal resistance.  The DC behavior of the cell in this regime is determined entirely by the summed impedances of the diode and the series resistance: 
\begin{equation}\label{eq:mSer} 
    V_{fb} = V_{di} + V_{Rser} = \textrm{ln}(I_{Out}/I_s)nk_B T + I_{Out} R_{ser} \ ,
\end{equation}
where $k_B$ is Boltzmann's constant, $T$ is the temperature of the silicon, $n$ is the diode quality factor (typically between 1 and 2)\cite{Jain05}, and $I_s$ is the diode saturation current.  We measured $R_{ser}$ for 37 solar cells by fitting Equation \ref{eq:mSer} to the portion of the unilluminated IV curve for which $V_{fb} \geq 0.5V$.  We let $R_{ser}$, $nk_B T$, and $I_s$ vary freely in the fit.  

In Figure \ref{fig:RShuntVsSeriesAndDarkIV}, we show the measured values of $R_{sh}$ and $R_{ser}$, along with contours for the $r$ figure of merit assuming a shorted load.  

After deciding to preserve those cells with the largest values of $r$ for future projects, we measured the full, unilluminated room-temperature IV curve of the C60 solar cell ``AX'' after affixing electrical leads by the process described in Section \ref{sec:prep}.  We show this curve in Figure \ref{fig:RShuntVsSeriesAndDarkIV}.  
To measure the ohmic behavior of the cell for small DC bias voltages, we fit a line to those points for which the magnitude of the applied voltage is below $100$mV (red points in the right panel of Figure \ref{fig:RShuntVsSeriesAndDarkIV}) and took the inverse of the fitted slope as the shunt resistance, $R_{sh}$. To measure the cell's series resistance, $R_{ser}$, we fit the IV curve for points with forward-bias voltages above 0.5V (blue points in the right panel of Figure \ref{fig:RShuntVsSeriesAndDarkIV}) using Equation \ref{eq:mSer}. 
These two slopes imply an $R_{sh}$ and $R_{ser}$ of $1010\Omega$ and $0.192\Omega$, respectively.  By fitting the current of the cell for applied voltages less than 3mV in magnitude, we measured the cell's intrinsic dark current, $I_B$, to be 2nA. 

Although these cells partially conform to the behavior predicted by the equivalent circuit of Figure \ref{fig:equivcircuit}, they are non-ohmic.   For example, cell AX has a nonlinear reverse current response to increasing back-bias voltages (negative voltages in the right panel of Figure \ref{fig:RShuntVsSeriesAndDarkIV}).  A back-bias voltage of 400mV drives 3.4 times more reverse current through the solar cell than the equivalent circuit predicts, with the disparity growing more extreme as the back-bias voltage grows. 

\subsubsection{A Hypothetical Method for Cancelling of DC Currents}

We anticipate using these devices in a mode where the light source of interest is modulated in order to distinguish calibration light from background illumination. 
The symmetrical ohmic behavior of these devices near zero applied bias voltage opens up the interesting possibility of applying an actively controlled bias voltage in order to drive a leakage current that cancels the sum of dark current and DC photocurrent from background illumination. This will reduce the Poisson noise at the modulation frequency, even if the cancellation is imperfect. Noise arising from fluctuations in the background level at the modulation frequency will remain a limitation. 

If the calibration light is modulated with a square wave, a fast A/D converter could monitor the net DC current when the light is in the off state and could change the applied bias to minimize this value.  An analog circuit could be designed to accomplish the same task. 

\subsection{Determining Non-Linearity of QE} \label{sec:linearityM} 

We define the nonlinearity of a solar cell as the dependence of its QE on the incident photon flux, $\phi$.  An ideal solar cell is perfectly linear, meaning its QE is entirely independent of $\phi$.
Our method for determining the nonlinearity of the C60 solar cells combines the simultaneous illumination of a reference and uncharacterized detector 
\cite{Emery2006} with the use of a lock-in amplifier \cite{hamadani2016non} (LI).  

We used a Stanford Research Systems (SRS) DS345 function generator to drive a set of Thorlabs LEDs, listed in Table \ref{tab:leds}, with a 13Hz 0-2V square wave.  We connected the LEDs to a 1.5mm diameter optical fiber, which illuminated one port of a Thorlabs 2" integrating sphere (IS) through a logarithmically varying neutral density (ND) filter and an iris.  A Thorlabs SM1PD2A photodiode (PD) occupied another IS port, and the solar cell stood several inches away from a third IS port.  The IS ports for the fiber, the PD, and the solar cell were mutually orthogonal.  The solar cell's collecting area was overfilled by the output of the IS.  BNC cables connected both the solar cell and the PD to a BNC switch box, the output of which we connected to an SRS SR570 current preamplifier.  We fed the output of the preamplifier into the input of a SRS SR2124 LI that was synchronized to the frequency of the function generator.  The BNC switch box toggled between the photodiode and solar cell signals, both subject to the same loading effects of the downstream instruments.  By moving the position of the variable ND filter with a translation stage, we changed the illumination intensity of the IS. 

\begin{table}[ht]
    \centering
    \caption{The Thorlabs LEDs${^*}$ used to study the linearity of the C60 solar cells.}
    \label{tab:leds}
    \begin{center} 
    \begin{tabular}{|c|c|c|}
    \hline
    LED ID${^*}$  & Nominal Wavelength (nm) & Nominal power through 1500$\mu$m dia. fiber (mW) $^{\dagger}$\\
    \hline
          M365FP1    &  365 & 300.0 \\ 
 M415F3 & 415 & 390.0 \\ 
 MINTF4    & 554 & 480.0 \\ 
 M625F2 & 625 & 320.0 \\ 
 \hline
    \end{tabular}
    {\raggedright  ${^*}$More information available at thorlabs.com/newgrouppage9.cfm?objectgroup\_id=5206 \par}
    {\raggedright  ${^{\dagger}}$ Inferred from reported power through a 200$\mu$m fiber assuming optical power scales with cross section area.   \par}
    \end{center} 
\end{table}

\begin{figure}[H]
\begin{center}
\begin{tabular}{c}
    \includegraphics[width=12cm]{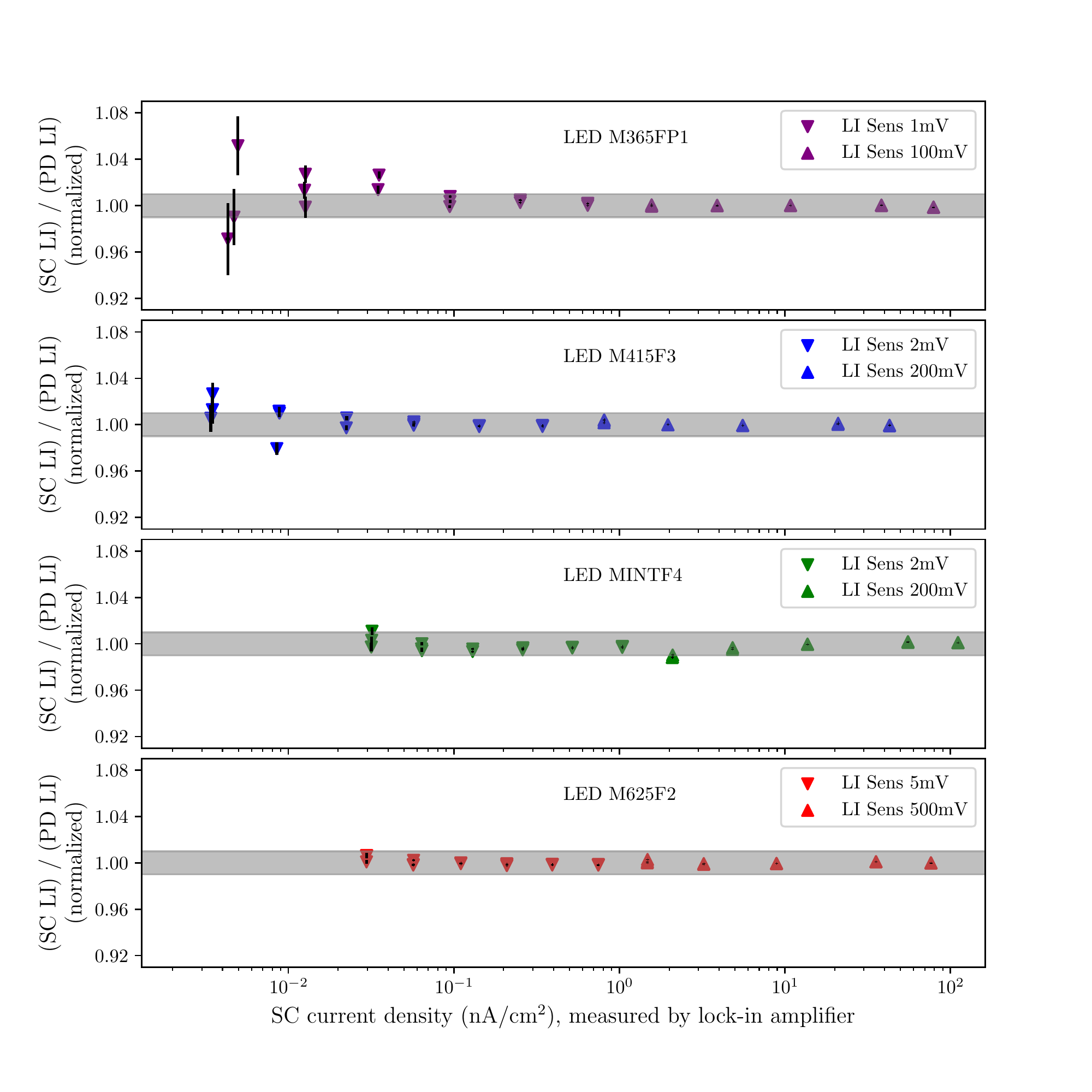}
\end{tabular}
\end{center}
    \caption{\label{fig:LInonlinearity} C60 solar cell (SC) nonlinearity measured using a lock-in amplifier (LI)  for the LEDs listed in Table \ref{tab:leds} measured relative to a Thorlabs SM1PD2A photodiode (PD).  The ratios have been normalized to unity, and the gray shadings show the $\pm$1\% linearity bounds.  Based on these results, the solar cell is linear to within $1\%$ for photocurrent densities between $\sim$100pA/cm$^2$ and $\sim$100nA/cm$^2$ for incident photon wavelengths between 365 and 625nm.} 
\end{figure}

In Figure \ref{fig:LInonlinearity}, we plot the normalized ratio of the solar cell and photodiode modulated currents as a function of solar cell current densities assuming the solar cell surface was illuminated homogeneously.  This ratio is constant to within 1\% over the range of monitored fluxes, suggesting that the C60 solar cells are linear to within 1\% for photocurrent densities between $
\sim$100pA/cm$^2$ and $\sim$100nA/cm$^2$, for the wavelengths tested.

\subsection{Rise Time/Frequency Response} \label{sec:frequency} 

We are interested in applications where the calibration light is modulated at some modest frequency to discriminate calibration light from ambient light. 
To determine a solar cell's viability as a calibrator for such a modulated light source, we must determine two semi-distinct frequency dependencies.  
\begin{enumerate} 
\item As the modulation frequency increases, more photocurrent will be discharged over the solar cell's shunt capacitance.  
\item If the effective QE of the detector diminishes as the modulation frequency of the light source increases, possibly due to non-negligible collection and diffusion times of the charge carriers \cite{Gudovskikh07, Morasero08}, the produced photocurrent will diminish and the output signal will drop. 
\end{enumerate} 
Both of these effects could reduce the amplifier signal as the frequency of modulation increases.     

\begin{figure}
\begin{center}
\begin{tabular}{c}
\includegraphics[height=6.0cm]{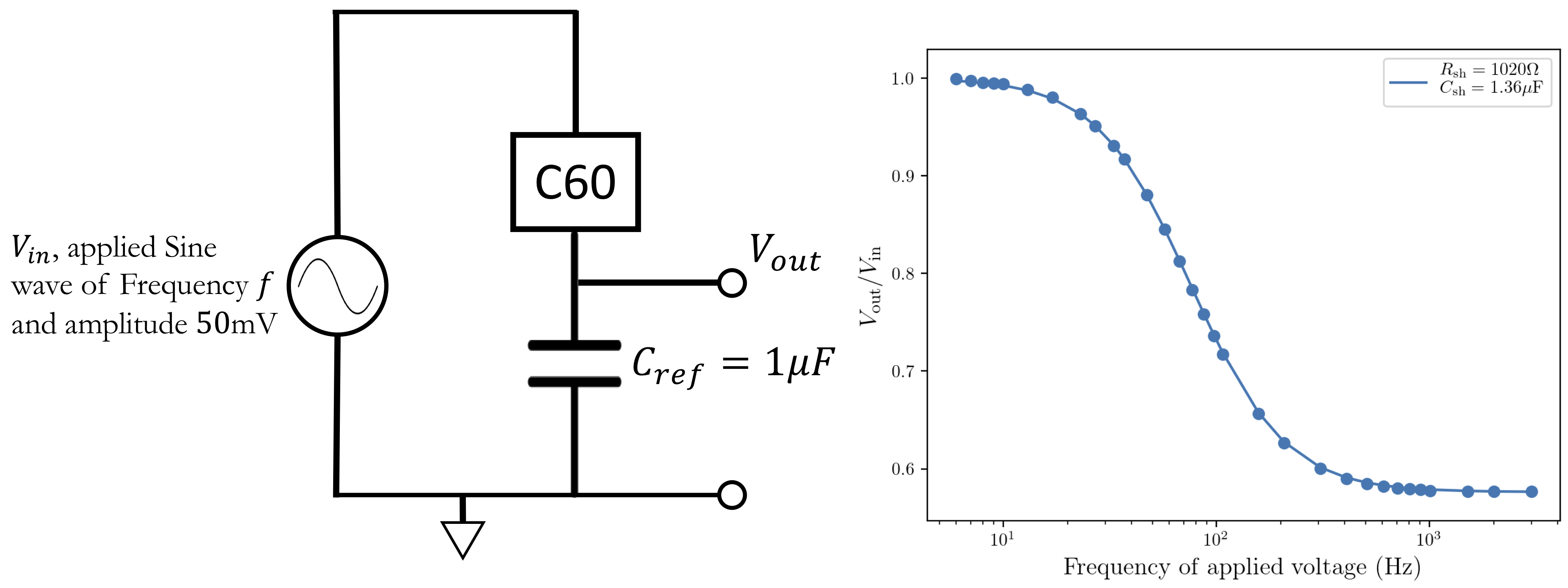}
\end{tabular}
\end{center}
    \caption{ \label{fig:capResults} Circuit configuration for measuring frequency dependence of C60's parallel impedance (left) and the results of the measurement (right).  We fit the measured voltage ratio values (points) to the voltage ratio predicted in Equation \ref{eq:capVoltRatio} with the extracted best-fit values of $R_{\mathrm{sh}} = 1020 \mathrm{\Omega}$ and $C_{\mathrm{sh}} = 1.36\mathrm{\mu F}$ (line).  The $\sim 1\%$ disparity between this measurement of $R_{sh}$ and the measurement of $R_{sh}$ derived from the unilluminated IV curve (Figure \ref{fig:RShuntVsSeriesAndDarkIV}) is likely due to the cell's non-ohmic IV response.}
 \end{figure}

We measured the shunt impedance of the solar cell, $Z_{SC}(f)$, by measuring the voltage across a series capacitance of $1\mu F$ given an applied 50mV peak-to-peak sinusoidal voltage of varying frequency.  The output voltage, $V_{\textrm{out}}$, is determined by $Z_{SC}(f)$ and by the impedance of the reference capacitor, $Z_{ref}(f) = -i (2 \pi f C_{ref}) ^ {-1}$, according to
\begin{equation} \label{eq:capVoltRatio} 
\frac{V_{\textrm{out}}(f)}{V_{\textrm{in}}(f)} = \frac{Z_{ref}(f)}{Z_{ref}(f) + Z_{SC}(f)} = \frac{1}{1 +  2 \pi i f C_{ref} Z_{SC}(f)} \ .
\end{equation} 
In Figure \ref{fig:capResults}, we show both the measurement scheme and the results of applying this measurement to solar cell ``AX''.  We found that the equivalent-circuit model of Figure \ref{fig:equivcircuit} effectively describes the shunt impedance of the solar cell up to an input frequency of at least $3 \mathrm{kHz}$ for voltages $\leq 50$mV.  

We repurposed the configuration used to measure the solar cell linearity (Section \ref{sec:linearityM}) to measure the solar cell's response to modulated light.  Our results are dominated by the gain bandwidth product of the current amplifier and by the upper frequency limit of the LED driver.  We determined that the solar cell has an acceptable time response out to at least 200Hz.

\subsection{Spatial Uniformity of QE} \label{sec:spatial} 
\begin{figure}
\begin{center}
\begin{tabular}{c}
\includegraphics[height=6.5cm]{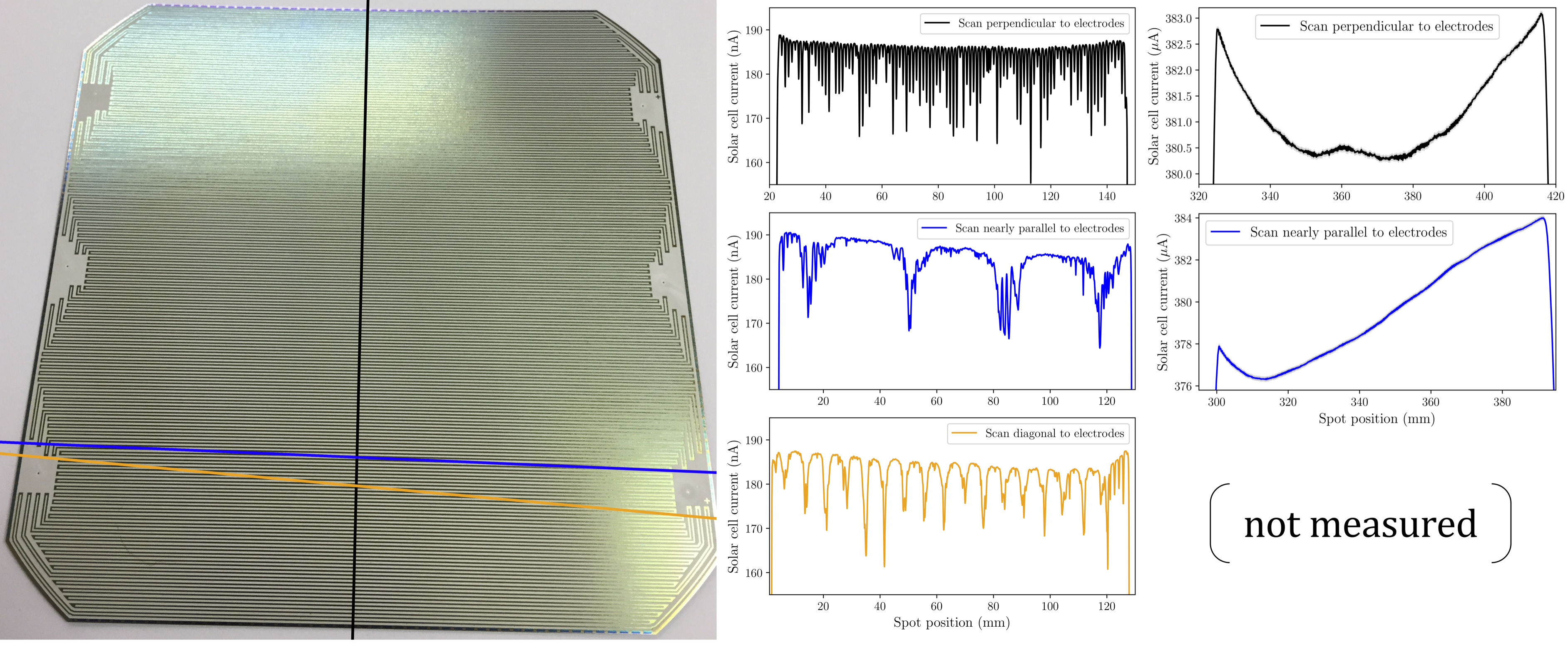}
    \end{tabular}
\end{center}
\caption{ \label{fig:spatial} C60 spatial response to a 150$\mu$m diameter spot (middle column) and to a 30mm diameter spot (right column) as the spot is moved perpendicular to the electrodes (black), nearly parallel to the electrodes (blue), and offset 10 degrees from parallel to the electrodes (orange).  For the small spot moving perpendicular to the electrodes, the SC signal is characterized by a series of sharp ``troughs'' in which the response drops by between 1\% and 17\%.  These troughs are 1.2mm apart.  The considerable structure within the troughs of the small spot's diagonal and parallel scans indicates that the solar cell response is highly nonuniform along the troughs.  Some $\leq 0.04\%$ variations at the $\mathcal{O}$(1mm) spatial scale are apparent in the solar cell response to both the the perpendicular and parallel scans of the 30mm spot.}
\end{figure}

The electrodes along the back of the C60 solar cells (see Figure \ref{fig:deviceimages}) produce a corrugated electric potential which may adversely effect the QE.  
To determine the strength of this effect, if it exists at all, we illuminated a 30$\mu$m pinhole with the M415F3 Thorlabs LED (see Table \ref{tab:leds}).  We used a pair of convex lenses to image the pinhole onto the surface of solar cell ``AX'', forming a 150$\mu$m diameter spot.  We used a translation stage to scan the pinhole image across the cell parallel to the electrodes, perpendicular to the electrodes, and diagonally with respect to the electrodes.  We show the results of these three scans in Figure \ref{fig:spatial}.  

The response of the solar cell as the $150\mu$m spot is moved parallel to the electrodes is characterized by a series of sharp ``troughs'' of varying depth.  Fourier analysis indicates that the spacing between these troughs is 1.2mm, $\sim2.5$ times larger than the spacing between adjacent electrodes of opposite polarity.  It is unclear if this spatial variation is caused by the corrugations in the electric field produced by the electrodes or due to some other regularly structured spatial inhomogeneity in the solar cell.  

In the parallel and diagonal scans, the spot obliquely traversed the electrodes and produced a series of much broader troughs in the measured current.  These broader troughs contain considerable substructure, suggesting that the QE response along the electrodes is not homogenous.  
 
  This spatial dependence of the cell's QE introduces a source of systematic error that will diminish as the cross section of the incident light grows.  To assess the effect of this systematic error on large optical beams, we scanned a 30mm diameter spot across the cell parallel and perpendicular to the electrodes.  The results of these scans are shown in Figure \ref{fig:spatial}.  Spatial variations in induced current on the scale of the electrode width are visible in both the perpendicular and parallel scan but are no bigger than $0.04\%$.  These scans also exhibit a $\mathcal{O}$(10-100mm) large-scale spatial inhomogeneity in the cell's QE that is as large as 4\%.  When using these devices as precision calibration tools, the location of the incident beam should be kept stable to minimize the systematic error introduced by such large-scale spatial inhomogeneities.  
  

\subsection{Statistical Uncertainty} \label{sec:noise} 
We are primarily concerned with the statistical uncertainty due to Johnson noise \cite{Nyquist28} in the resistances  and due to shot noise in the photocurrent, $I_L$.  at the $\mathcal{O}$(1mm) spatial scale
If we assume that the photocurrent is substantially larger than both the bias current and the current driven over the shunt resistance due to the load's bias voltage, we can estimate the statistical uncertainty in the load current, $\sigma_{I, Out}$, by adding the shot noise and the Johnson noise in quadrature:
\begin{equation}  \label{eq:currentSNR} 
SNR = \frac{<I_{Out}>}{\sigma_{I,Out}} = \frac{<I_{L}>}{\sqrt{2<I_L> e \Delta f + 4 k_B T \Delta f (1+1/r) R_{sh}^{-1}} } \ ,
\end{equation} 
where $<\cdot>$ denotes the expectation value, $e$ is the electron charge, $\Delta f$ is the monitored bandwidth, $k_B$ is Boltzmann's Constant, $T$ is the temperature of the solar cell, $R_{sh}$ is the shunt resistance, and $r$ is the resistor ratio figure of merit defined in Section \ref{sec:dark}.  For a solar cell with $r>>1$ and $R_{sh} = 500\Omega$ operating at $T=300$K, the expected contributions from shot noise and Johnson noise are equal when $I_L \simeq 100 \mu$A.  Poisson-noise-limited performance requires $I_{L} > 100 \mu$A. 

\section{Conclusions and Next Steps}\label{sec:conclusion} 

We have measured the electrical properties, linearity, frequency response, spatial uniformity, and noise characteristics of C60 back-contact monochrystalline solar cells.  We undertook this exploration in the hopes that these cells might serve as precise photometric calibrators for the LSST's Collimated Beam Projector (CBP) and for other large area light sources.  Our results are promising.  

In Section \ref{sec:dark}, we measured the electrical properties of unilluminated solar cells assuming the equivalent circuit model of Figure \ref{fig:equivcircuit}.  We found that typical cells have shunt resistances $\mathcal{O}(100-1000\Omega)$ and series resistances $\mathcal{O}(0.01-0.1\Omega)$.  The modest shunt resistances mean that even 0.1mV of applied bias voltage can produce $\mathcal{O}(0.1-1\mu\mathrm{A})$ of dark current.  When using these solar cells as precision calibrators, steps should be taken to minimize inadvertently applied bias voltages by, e.g., non-ideal current amplifiers.  We also measured an extensive IV curve for a single solar cell, and found a nonlinear current response to back bias voltages, presumably due to quantum tunneling.  

In Section \ref{sec:linearityM}, we used a lock-in amplifier and a set of LEDs to measure the dependence of a cell's QE on the total incident flux (it's `linearity').  The cell showed less than 1\% variation in QE for photocurrent densities between 0.01 nA cm$^{-2}$ and 100 nA cm$^{-2}$.  

 In Section \ref{sec:frequency}, we determined that the cell's frequency response to a modulated light source is nearly unity for modulation frequencies less than 200Hz.  Additional study is needed before they can be used for higher modulation frequencies.  

In Section \ref{sec:spatial}, we found that the solar cell QE varied by as much as 17\% as we moved a 150$\mu$m diameter spot perpendicular to the cell's electrodes.  We suspect that this QE variation results from corrugations in the cell's electrical potential, though the apparent mismatch between the electrode spacing (0.46mm) and the spacing of the QE troughs (1.2mm) is something that we do not yet understand.  This spatial variation in QE constitutes a source of systematic error that will diminish as the cross section of the incident beam grows.  For a beam of diameter $\geq$30mm, the systematic error in QE resulting from this effect is $\leq$0.04\%.  

In Section \ref{sec:noise}, we discussed some sources of statistical error.  
The statistical fluctuations in our measurements will be dominated by shot noise for photocurrents $\gtrsim 100 \mu$A.  

As we described in Section \ref{sec:dark}, the flow of current over the resistance of attached electrical leads will drive a smaller current through a cell's modest shunt resistance.  Variation with time ore termperature in these resistances will produce a commensurate time-dependence in the cell's effective QE, introducing a systematic error.  We believe this effect constitutes the largest obstacle to employing these cells as sub-percent photometric calibrators.  We are also optimistic that this obstacle can be surmounted by minimizing the resistance of the leads and of the connections to the current amplifier, by selecting those cells with the largest shunt resistances, and by carefully controlling systematic voltage and resistance drifts.  

Our results and the vendor's reported quantum efficiency\cite{C60Brochure} suggest that back-contact monochrystalline solar cells could serve as precision photometric calibrators for large beams.  

\section{Acknowledgments}

We thank the US Department of Energy and the Gordon and Betty Moore Foundation for their support of our LSST precision calibration efforts, under DOE grant DE-SC0007881 and award GBMF7432 respectively. This material is also based partially upon work supported by the National Science Foundation Graduate Research Fellowship Program under Grant No. DGE1745303.  Any opinions, findings, and conclusions or recommendations expressed in this material are those of the authors and do not necessarily reflect the views of the National Science Foundation.


\bibliography{SolarCells.bib} 
\bibliographystyle{spiejour}   


\vspace{2ex}\noindent\textbf{Sasha Brownsberger} is a graduate student seeking his PhD at Harvard University, working with Professor Christopher Stubbs.  He received his BS in Physics and Mathematics from Stanford University in 2010.  His current research interests include the search for astronomical probes to study the expansion history of the universe and the development of the next generation of astronomical detectors. 

\vspace{2ex}\noindent\textbf{Nicholas Mondrik} is a PhD student at Harvard University working with Professor Christopher Stubbs.  His current research interests include calibration of large-scale photometric surveys, development of astronomical instrumentation and calibration hardware, and the study of M-dwarfs as exoplanet hosts.  

\vspace{2ex}\noindent\textbf{Christopher Stubbs} is a Professor of Physics and of Astronomy at Harvard University.  His BS in Physics was awarded by the University of Virginia, and he was granted a PhD in Physics by the University of Washington.   

\listoffigures
\listoftables

\end{spacing}
\end{document}